**RESEARCH ARTICLE**

**Structural studies of the silica sol-gel glasses doped with copper selenide nanoparticles with plasmonic resonance absorption**

V.S. Gurin
*Research Institute for Physical Chemical Problems, Belarusian State University, Minsk, Belarus*
*gurin@bsu.by*

A. A. Alexeenko
*Gomel State Technical University, Gomel, Belarus*

## Abstract

**Background:** Semiconductor-doped glasses are treated actively through many years and continue to be of great interest because challenged features of nanosized semiconductors of various chemical nature. Copper chalcogenides have discovered the plasmonic properties in line with quantum confinement effects specific for major of semiconductor nanoparticles.

**Objective:** The aim of this work is to study structural and optical features of the sol-gel derived silica glasses with copper selenide nanoparticles demonstrating appearance of the plasmonic light absorption in the near IR range.

**Method:** The samples under study were fabricated through an original sol-gel technique realizing the simultaneous synthesis of copper selenide and sintering of mesoporous silica. The copper selenide glasses were characterized with X-ray diffraction (XRD), transmission electron microscopy (TEM) and optical absorption spectroscopy.

**Results:** Formation of nanocrystalline $Cu_{2-x}Se$ particles of the size range from tens nm through 100-150 nm is established with XRD and TEM techniques. The principal optical properties are presented by the featured absorption in the visible and near-IR ranges. $E_g$ was evaluated for the direct transitions in the range of 2.10-2.36 eV. The plasmonic resonance in the nanoparticles due to increased carrier concentration originated by intrinsic defectness of $Cu_{2-x}Se$ nanoparticles with variable stoichiometry. Its energy can be controlled by Cu/Se ratio in the synthesis procedure.

**Conclusion:** The silica sol-gel glasses with copper selenide nanoparticles were fabricated, characterized by XRD and TEM methods, and their optical absorbance spectra were investigated. The principal optical properties are presented by the featured absorption in the visible and near-IR ranges: the step-like proper absorption of the semiconductor particles characterized by $E_g$ and the intense near-IR band is associated with the localized plasmonic resonance in the nanoparticles due to increased carrier concentration.

# 1. Introduction

A history of semiconductor nanoparticles has been introduced by investigation of such 'classical' compounds as II-VI and I-VII compounds (e.g., CdS, ZnS, CuCl) [1-3] and their size-dependent features have been discovered. They possessed comparatively simple band structure, and basically, had direct band gaps. Novadays, advanced nanochemical synthetic methods allow to fabricate much more wide circle of binary, ternary and more complicated semiconductor nanoparticles and clusters, e.g. [4-7]. The nanoparticles in various media may be fabricated: colloids, crystals, ordered mesophases, amorphous materials like glasses, etc, and their properties depend strongly not on size and intrinsic particle features, but also on the environmental factors due to possible interactions between nanoparticles and surrounding medium.

Among many semiconductor compounds that actively studied last decades at the nanoscale, copper chalcogenides (and also chalcogenides of another transition metals, but CuX appeared to be pioneric) [8-11]. They reveal unique properties combining both quantum confinement effects and the plasmonic resonance phenomenon. The plasmonic features occur due to elevated free carrier concentration (electrons or holes, depending type of compounds). In particular, copper selenides possess variable composition and non-stoichiometry and include a number of stable and metastable phases for $1 \leq x \leq 2$, e.g. in $Cu_{2-x}Se$ [12,13]. The value of x strongly affects the carrier concentration and all other properties. In the case of nanoscale copper selenides, more features are expectable due to both confinement effects, size-dependent geometry modifications of particles, surface reconstruction, etc. In optics, new absorption bands and IR luminescence discovered and associated with the plasmonic resonance due to intrinsic defectness that provides creation of free carriers. However, advantage of the plasmons in such semiconductors is the variable carrier concentration resulting in efficient control of the spectral range of the optical response (basically, the near-IR range) [14-16].

The particles localized in glass matrices are of special attention forming the stable system for potential applications as active non-linear optical media [17,18]. An investigation of the state of nanoparticles within glassy materials is required for proper control of their composition, size, interaction with the matrix and environment. Some basic structural features of these materials have been found in a series of previous publications [19-24]. However, there was no systematic analysis to date for all key results on structural features, in particular, the information obtained by X-ray diffraction (XRD) and transmission electron microscopy

(TEM). These routine techniques of structural analysis of solids meet some issues for the semiconductor-doped glasses because rather small relative content of active semiconductor in glass and amorphous nature of the matrix. Any extraction of the particles with a glass destroy can essentially change information on the particle state. Therefore, in the present work, we consider the results of XRD and TEM study of the sol-gel derived glasses with copper selenide nanoparticles those present additional data for our previous findings regarding their structure and optical features.

## 2. Experimental: fabrication of samples and methods

The fabrication procedure of the silica sol-gel glasses in this work is based on the sol-gel technique with hydrolysis of organic ethers of siliceous acid that has been developed ago and widely used for preparation of very versatile circle of materials: glasses, ceramics, films, etc. [25-27]. The conventional preparation workflow has been modified by incorporation of precursors for formatin of metallic copper and copper compounds (copper oxides and copper selenides) together with prevention of volume contraction of the glass samples under drying and sintering to keep good optical quality of final optical glasses. The following main steps are in this procedure:

(i) Preparation of precursor sol mixing alcohol-aqueous solution of tetraethoxysilane (TEOS) at the molar relation $TEOS/H_2O = 1/8$ with an acid catalysts (HCl), molar ratio acid/TEOS = 1/50.

(ii) Addition of aerosil ($SiO_2$ powder with the grain size about 20 nm) to the sol to avoid a volume contraction under further gelation and drying.

(iii) The gelation step due to pH increase up to 6-6.5 by addition of ammonia solution. To get samples of definite shape the sols were poured into polysterene containers, and gelation proceeded for 24 h.

(iv) Drying the gels at $60^{o}C$ followed by heating up to $600-1000^{0}C$ during 2 h to remove sorbed water and organic residues. The maximum temperature of this treatment manages properties of xerogels (density, porosity, amount of remnant hydroxyl groups, etc).

(v) The doping with copper was conducted through the impregnation of porous xerogels in alcoholic $Cu(NO_3)_2$ solution during 8 h followed by drying in air. Copper oxide (CuO) was produced at the heating($600^{o}C$) in air of the samples after this step.

(vi) The chemical transformation of copper oxide into metallic copper nanoparticles through

the heating in hydrogen flow.

(v) Production of copper selenide nanoparticles through interaction of metallic copper with selenium vapour. This step was realized within the sealed ampoules under elevated temperatures. The partial pressure of Se is controlled by the amount of elemental Se introduced into this reactor.

(vi) Annealing of the silica xerogels with copper selenide formed within it by the controlled heating regime up to $1200^0C$. This step resulted in production of transparent glassy samples of appropriate optical quality and good mechanical strength.

TEM study of the glass samples with nanoparticles was done through the "replica with extraction" method were used in the case of medium resolution electron microscope. A thin carbon film (10-20 nm) was evaporated onto a freshly etched surface of the samples followed by the carbon film detaching in water and transfer it to a TEM grid. This procedure is rather mild to expect any deteriorating of particles and had the deficiency that a part of particles can be not extracted, but the correspondence of the data obtained with the direct method for selected samples supports its adequacy for information on the particle state.

XRD measurements were performed for the thin glass samples without their destroy with DRON-3 and HZG-3 diffractometers using $CoK_\alpha$ and $CuK_\alpha$ radiations and the Bragg-Bretano geometry.

To get optical absorption spectra of the glasses they were cut down to the thickness about 0.3 mm and polished for good optical quality. The absorption spectra were recorded with a CARY 17D device throughout the UV/Vis/near-IR range. Undoped glasses, i.e the blank silica matrix, was fabricated in similar manner of heat treatment protocols. They had very low light absorption in this range, hence no special subtraction procedure was done for the spectra presented. XRD data evidenced no any new solid phase in them besides typical pattern of the amorphous silica [28]). Characteristics of the samples are collected in Table 1.

## 3. Results and Discussion

XRD study in this work was performed to get information on the phase composition of the whole "particle-matrix" system and properly the copper selenide nanoparticles embedded.

Table 1. The samples of sol-gel glasses doped with copper selenide, characteristics of their composition and the value of optical band gap evaluated from the absorption measurements (Fig. 5).

| Sample code | Cu content in the precursor solution, mmol/l | Cu content in samples calculated as $Cu/SiO_2$ atomic ratio | Pressure of Se under $Cu_xSe$ synthesis, atm | $E_g$, eV | $E_{max}$, EV |
|---|---|---|---|---|---|
| M387 | 3 | 0.13 | 0.1 | 2.36 | 1.02 |
| M388 | 3 | 0.13 | 0.2 | 2.14 | 1.08 |
| M394 | 20 | 0.40 | 0.1 | 2.14 | 0.94 |
| M401 | 6 | 0.13 | 1 | 2.12 | 1.10 |
| M402 | 20 | 0.40 | 1 | 2.10 | 1.19 |

However, the particles are not detectable in all cases of their location within the glass. The low their concentration and the background effect of the amorphous silica prevent appearance of good diffraction patterns. In combination with TEM data the presence of nanoparticles is evidenced unambiguously in the doped glasses. Undoped ones are amorphous silica. Its detailed investigation is out the framework of this publication.

Figs. 1,2 present the XRD patterns for a series of the glass samples with copper selenide nanoparticles fabricated under various precursor concentrations. They evidence formation of the nanocrystaline $Cu_xSe$ phase that can be assigned to the cubic structure, space group F4(-)3m (No.216), the database PCPDFWIN, entry JCPDS 06-680. For the sample M402 (best appearance in the set of XRD data) the unit cell parameter can be evaluated considerable exactly, a = 5.728 . This value is very close to the reference one (JCPDS 06-680) a = 5.739 . Therefore, for this sample the structure of $Cu_{2-x}Se$ nanoparticles fits practically with the bulk copper selenide, however, the stoichiometry is slightly variable, $Cu_{2-x}Se$, also similar to the bulk state of this compound.

The first set of the XRD patterns demonstrates that for the samples with relative low concentration of copper precursor (and correspondingly, copper selenide produced) an appearance of the cubic $Cu_{2-x}Se$ (as noted above) may be decoded, although the peaks are very weak. Exact evaluation of the unit cell parameters is troubled, while an approximate value is consistent with the above for the best XRD pattern.

As the whole, XRD pattern of these sol-gel silica glasses doped with copper selenide nanoparticles (Fig. 1,2) are produced due to scattering on amorphous silica matrix and the nanocrystalline phase of nanoparticles of a low volume concentration. Therefore, in the typical pattern like given in Fig. 1a the broad and intense halos at 2θ in the range of 20-30 deg dominates. It is known to be originated from the scattering at disordered glassy medium. A proper analysis deriving atomic pair distribution function may be applied to evaluate more information on its structure [28], however, this is out the framework of the present paper. A series of diffraction peaks due to the Bragg scattering at copper selenide particles enter the range of more 2θ values that fortunately provide the feasibility to establish the presence of $Cu_{2-x}Se$ nanoparticles.

The electron microscopic images given in Fig. 3 (medium resolution TEM technique) clearly demonstrate presense of nanoparticles for the selected set of samples. The picture is similar for all the glasses under study, the difference occurs basically in concentration of particles and their size. The size range is tens of nanometers (averaged, ~10-50 nm), the shape is near spherical. Notably, the particles are localized in the specific areas (cavities) within the deformed glass. This means that the process of particle formation has the noticeable effect upon the amorphous silica matrix than can be provided by the different behavior of copper selenide and silica under cooling down from the maximum annealing temperature (1200°C), while the melting point of copper selenide is 1113°C [29]. Remarkably, the particles are slightly greater for the sample with higher copper content (M402) while the concentration observed by the particles occurrence is not very different for the samples presented.

Fig. 4 presents a series of absorption spectra of all samples. First of all, one should be noticed, that all they are untypical for semiconductor nanoparticles like II-VI and IV-VI chalcogenides (e.g. CdSe, ZnSe, PbSe, etc). The latter for nanoscale particles are modified as compared from the bulk counterparts by the appearance of excitonic maxima those correspond to a series of resolvable excitations, unresolved excitons provide just the band gap feature in the higher-energy part of spectra. Here, the combination of the step-like feature in the visible range (0.4-0.6 mm) and the broad intense band in the near IR range.(0.8-1.6 mm).

The step-like part of these spectra corresponds to the proper absorption edge of semiconductor. Assuming the direct-band character for copper selenide, the values of $E_g$ were derived (Fig. 5). The conventional Tauc formula was used for this procedure ploting $(\alpha E)^2$ vs E, where E is photon energy and $\alpha$ stands for absorption coefficient. There are linear parts in these dependencies those result the band gap values collected in Table 2. Surprisingly, they appear to be almost same, besides, the value $E_g$=2.36 eV for one sample which was fabricated with minimum amount of Cu and Se precursors (the sample M387). Meanwhile, the other three samples for which the ratio Cu/Se varies rather considerably (Table 2), $E_g$ shows the less value (2,12-2,14 эВ). Such little variance in $E_g$ values for the set of samples under study seems to be rather unexpected as far as the stoichiometry of $Cu_{2-x}Se$ particles produced can depend strongly on the Cu/Se ratio which directly influence semiconductor properties of copper selenide. This occurs evidently in the case of bulk $Cu_{2-x}Se$, however, the conditions of the nanoparticles formation in the procedure used for $Cu_{2-x}Se$ in sol-gel silica matrix can change the situation. Possible reason of almost constant $E_g$ can be suggestion that the process of copper selenide formation within the closed vessel and mesoporous silica matrix at $1200^o$C and enough long time realize an equilibrium phase and excess of precursors remained to be unreacted. According to the XRD data, the main phase is berzelianite $Cu_{2-x}Se$, $0 < x < 0.25$. Within this narrow interval of $x$, the minimum amounts of Cu and Se precursors correspond to the higher value of $E_g$.

The near-IR band possesses rather similar profile for all set of samples and its maximum position is slightly varied ($E_{max}$, Table 1). There is the tendency of shifting $E_{max}$ to the higher energies for the samples fabricated under higher selenium pressure. Probably, the higher $P_{Se}$ can provide formation of $Cu_{2-x}Se$ with more $x$ analogously to the above trend in change of $E_g$. Therefore, the changes of the precursor amounts are the working method to control principal parameters of the optical features of the glasses under study. Meanwhile, the range of these variations is not too wide for the sample set presented here.

The nature of this band can be associated with appearance of the plasmonic resonance in these nanoparticles in which an increased number of charge carriers exist due to intrinsic defects ruled by the variable stoichiometry [12-16]. Within the framework of the simple Drude approach (applicable for electronic gas in metals and more general free carriers in semiconductors) the plasmonic resonance frequency is described by the relation

$$\omega_p^2 = e^2 N / m \varepsilon_0, \qquad (1)$$

where N is carrier concentration, e is elementary charge, m does the effective mass of carriers, and $\varepsilon_0$ denotes the vacuum dielectric constant. Thus, $\omega_p$ is proportional to square root of N. For N values in the range of $10^{20}$-$10^{21}$ cm$^{-3}$ the plasmonic resonance enters the near-IR spectral range. No direct information on the carrier concentration is available within the framework of the present work, however for $Cu_{2-x}Se$ nanoparticles synthesized in the colloidal state [30] one can find the carrier concentration estimated in this range, and the plasmonic properties are similar with our present results.

## 4. Conclusion

In this work, the silica sol-gel glasses with copper selenide nanoparticles were fabricated, characterized by XRD and TEM methods, and their optical absorbance spectra were investigated. XRD study established the formation of nanocrystalline $Cu_{2-x}Se$ particles. The size range of the nanoparticles according to TEM data enters tens of nm increasing to 100-150 nm for the high precursor concentrations. The principal optical properties are presented by the featured absorption in the visible and near-IR ranges. The step-like part corresponds to the proper absorption of the semiconductor particles characterized by $E_g$ evaluated for the direct transitions in the range of 2.10-2.36 eV for the set of samples with $Cu_{2-x}Se$ nanoparticles under study. The intense near-IR band is associated with the localized plasmonic resonance in the nanoparticles due to increased carrier concentration. The latter is provided by intrinsic defectness of $Cu_{2-x}Se$ nanoparticles with variable stoichiometry. The lower plasmonic resonance energy (red shift) occurs for the higher Cu/Se ratio that may be controlled through the synthesis procedure.

This new example of semiconductor plasmonics is of interest for elaboration of novel optical devices in which size effects of the nanoscale semiconductors are combined with the strong optical response of free charge carriers tuned by the particle chemical composition. It is noticeable, that the plasmonic particles in glass possess many advantages over the intensive treated colloidal-synthesized counterparts because more stability of particles within the matrix.

FUNDING

The work was performed under support of the State Program of Scientific Investigations of Belarus "Nanostructure" (2021-2025).

CONFLICT OF INTEREST

Declared no conflicts

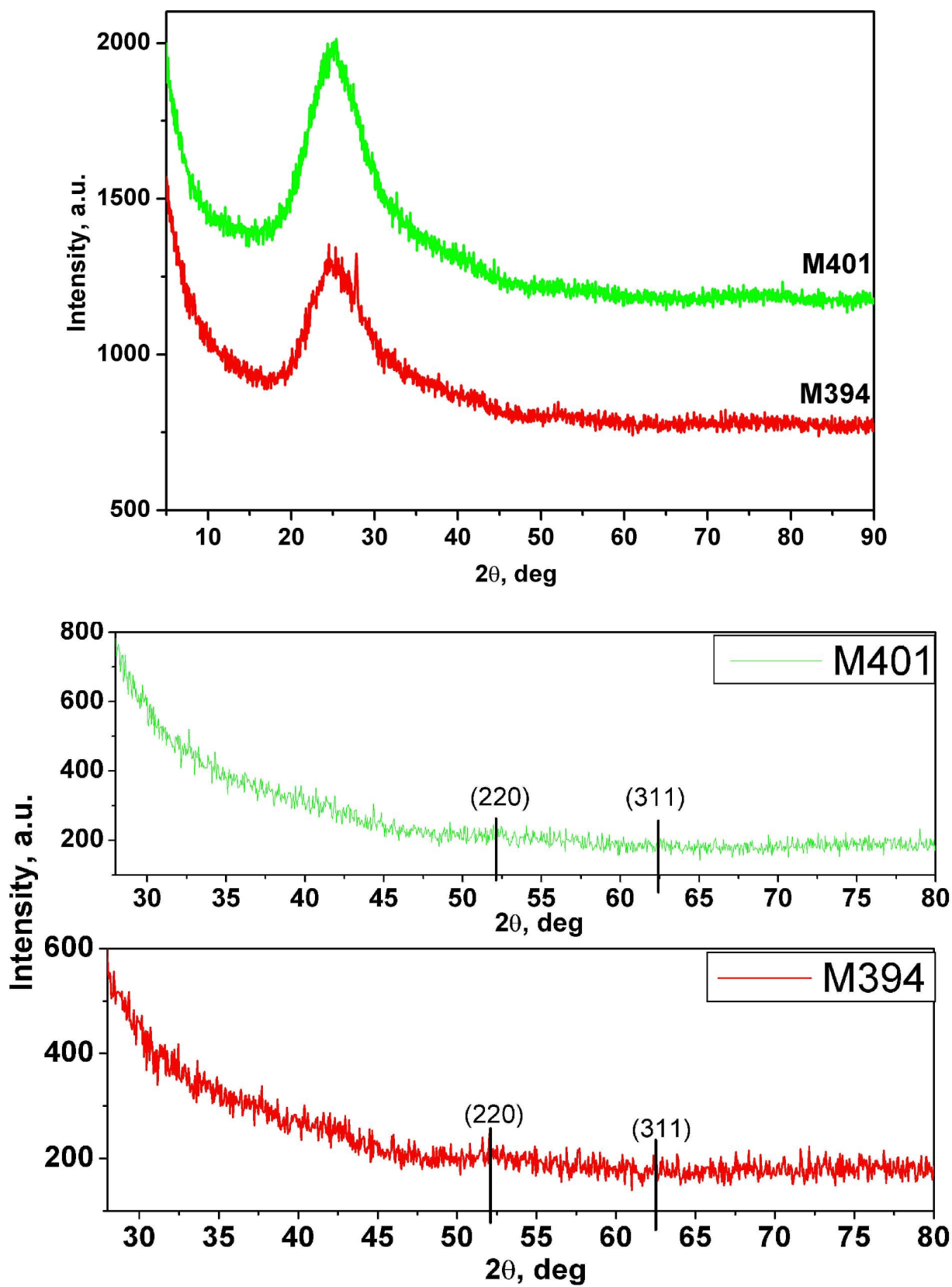

Fig. 1. XRD patterns of a series of glasses with copper selenide nanoparticles fabricated using various component contents (Table 1). Upper plot – the full range of diffraction angles, the three samples together. Lower plot – the range of angles to discover the weak peaks labeled correspondingly. $CoK_\alpha$ radiation was used for the measurements.

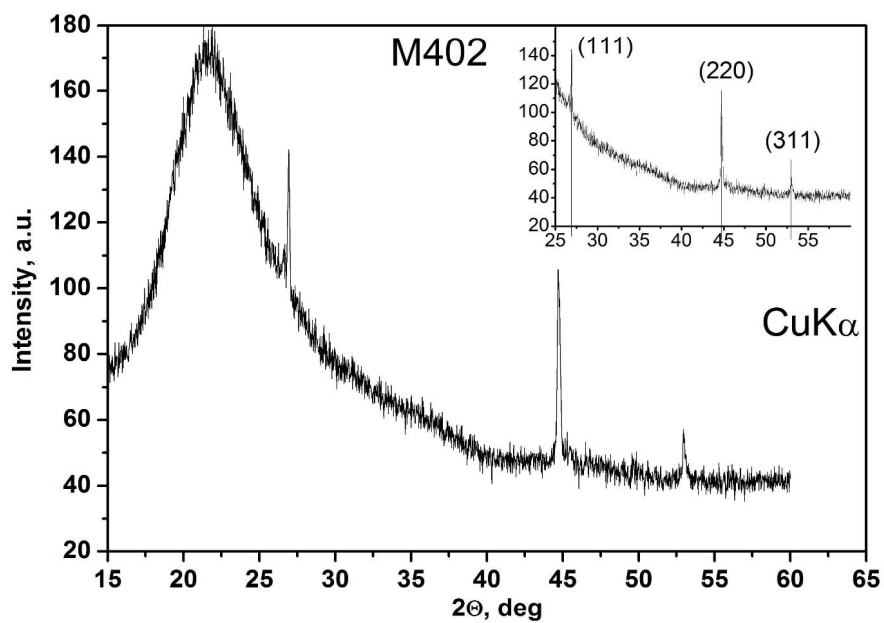

Fig. 2. XRD patterns of the sample M402 with high concentration of copper precursor (Table 1). CuKα radiation was used. Insert shows the shortened range of diffraction angles with the peaks decoded as $Cu_{2-x}Se$ berzelianite phase, JCPDS 6-680. See the body text for the structure refinement results.

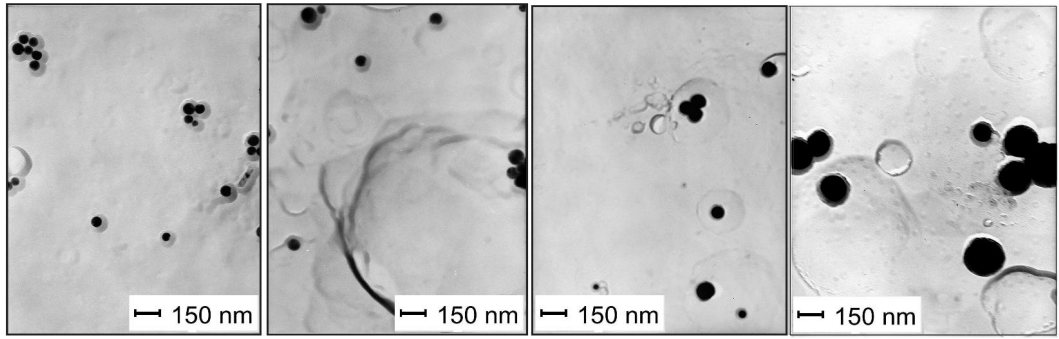

Fig 3. TEM micrographs of the sol-gel glasses with copper selenide nanoparticles.
The samples are (from left to right): M387; M388; M401, M402, the labels in Table 1.

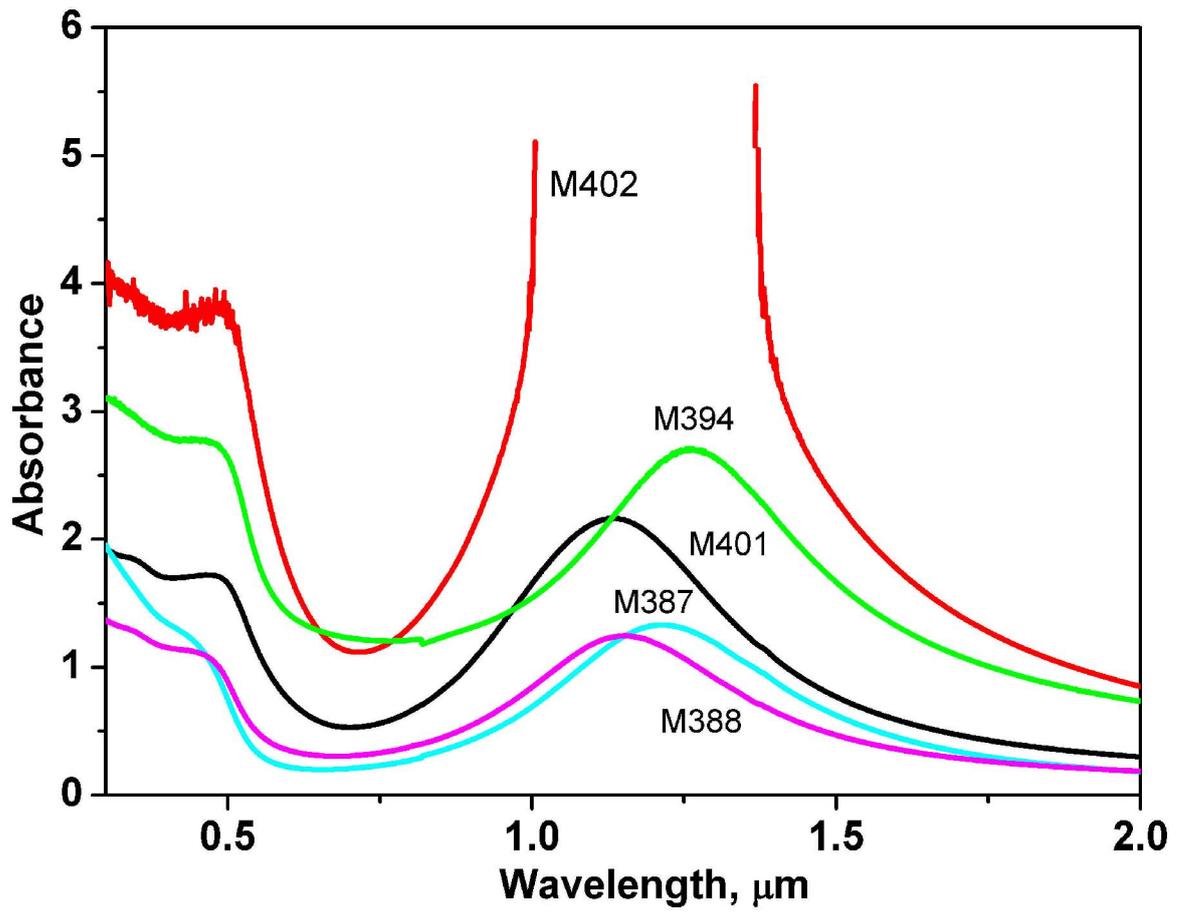

Fig. 4. Absorption spectra of a series of glasses under study. The corresponding sample
labels are described in Table 1.

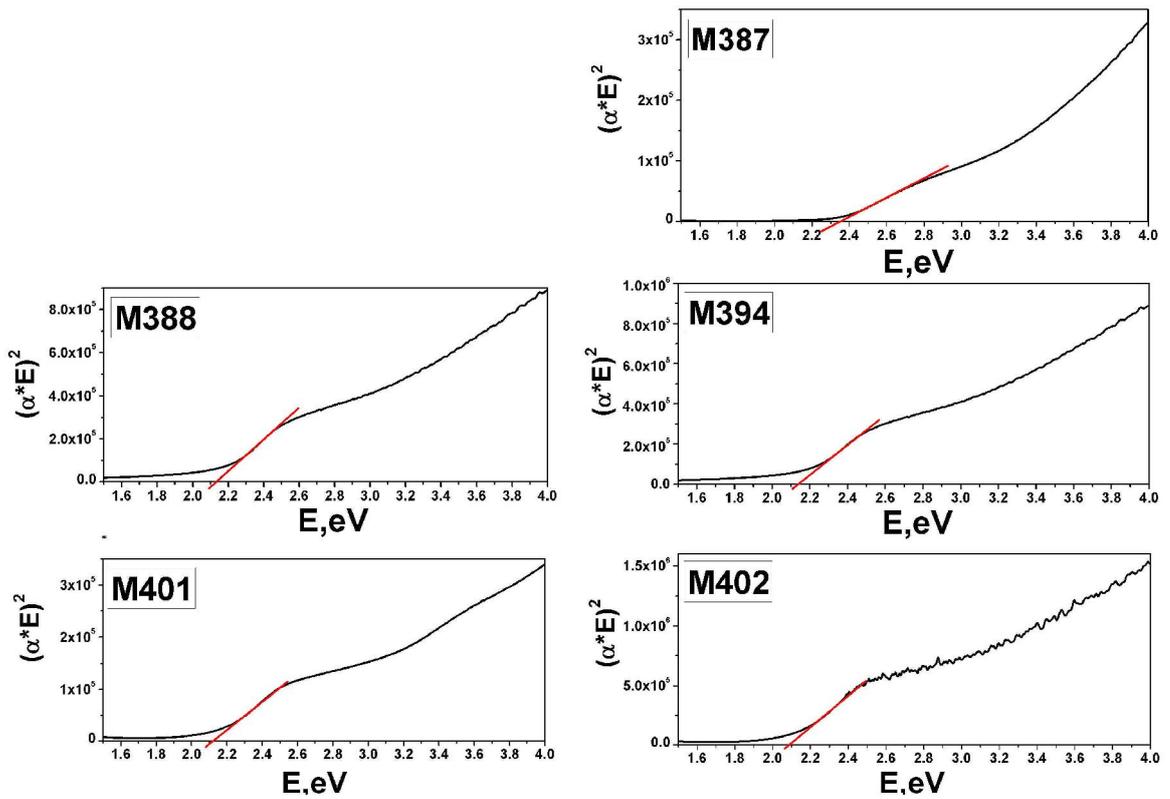

Fig. 5. Evaluation of $E_g$ from the absorption spectra for a series of glasses under study (Table1) through the plotting $(\alpha E)^2$ $(cm^{-2}eV^2)$ *vs* E (eV), where E is photon energy and $\alpha$ stands the absorption coefficient of the corresponding sample.